\documentstyle[preprint,aps]{revtex}                  
\begin{document}
\pagestyle{empty}                                      
\preprint{
\font\fortssbx=cmssbx10 scaled \magstep2
\hbox to \hsize{
\hbox{\hskip.2cm  hep-ph/9512239}
\hfill$\raise 1cm\vtop{\hbox{CERN-TH/95--326}
                \hbox{FISIST/14-95/CFIF}
                \hbox{NTUTH-95-11}\hbox{}}$}
}
\draft
\vfill
\title{
$R_b$ -- $R_c$ Crisis and New Physics
}
\vfill
\author{Gautam Bhattacharyya}
\address{
Theory Division, CERN, CH-1211 Geneva 23, Switzerland
}
\author{Gustavo C. Branco}
\address{
Centro de F\' \i sica das Interac\c c\~ oes Fundamentais,
Instituto Superior T\' ecnico - Dep. de F\' \i sica,
Avenida Rovisco Pais P - 1096 Lisboa Codex - Portugal
}
\author{Wei-Shu Hou}
\address{
Department of Physics, National Taiwan University,
Taipei, Taiwan 10764, R.O.C.
}
\date{\today}
%
%
\vfill
\maketitle
\begin{abstract}
The experimental values of $R_b$ and $R_c$ are the only data
which do not seem to agree with Standard Model predictions.
Although it is still premature to draw any definite conclusions,
it is timely to look for new physics
which could explain the excess in $R_b$ and deficit in $R_c$.
We investigate this problem in a simple extension of the Standard Model,
where a charge $+2/3$ isosinglet quark is added to the standard spectrum.
Upon the further introduction of an extra scalar doublet,
one finds a solution with interesting consequences.
\end{abstract}
%
%
\pacs{PACS numbers:
}
%
%
\pagestyle{plain}

Precision electroweak tests at the $Z^0$ resonance provide
impressive confirmation of the Standard Model (SM),
allowing for the extraction of $m_t$ via global fits to electroweak data,
which  is in good agreement with direct measurements at
the Fermilab Tevatron. Recently, however,
the so-called ``$R_b - R_c$" crisis has been reported \cite{lep},
namely,
an excess in $R_b \equiv \Gamma_b/\Gamma_{\rm had}$
and deficit in $R_c \equiv \Gamma_c/\Gamma_{\rm had}$,
with respect to their SM predictions.
{}From a multiparameter electroweak fit,
the latest results are
\begin{equation}
R_b^{\rm exp}  =  0.2219 \pm 0.0017, \ \ \ \
R_c^{\rm exp}  =  0.1543 \pm 0.0074,
\end{equation}
while, for $m_t = 180$ GeV,
$R_b^{\rm SM}  =  0.2156$, and
$R_c^{\rm SM}  =  0.172$.
Thus, with an experimental accuracy of $\sim 0.8\%$ ($5\%$),
the SM expectation for $R_b$ ($R_c$) lies $3.7 \sigma$ ($2.5 \sigma$)
below (above) the experimental result.
Of growing concern is that,
while the discrepancy had existed previously \cite{glas},
it became more acute after inclusion of 1994 data.
It should be noted that the measurements of $R_b$ and $R_c$
are rather correlated $(- 0.35)$,
therefore an improvement in the measurement of one
will also reflect on the measurement of the other.

It may be premature to use these measurements
to draw any definite conclusions.
Indeed, more data or better analysis methods
might bring $R_b$ and $R_c$ back at par with their SM predictions.
On the other hand, there is also the more exciting possibility that,
with time, $R_b$ might remain above the SM prediction and
$R_c$ below it, thus hinting at physics beyond the SM.
It should be kept in mind, however,
that the measurement of the total hadronic width
through the variable $R_{\ell} = \Gamma_{\rm had}/\Gamma_{\ell}$
is rather consistent with the SM.
One may therefore ask:
What sort of new physics can shift $R_b$ and $R_c$ in the
{\it right directions},
while keeping other observables consistent with
present experimental values?

In this paper we first explore the case
where $\Gamma_c$ is reduced
while $\Gamma_b$ is not changed.
This shifts $R_b$ and $R_c$ in the right directions.
A viable extension of the SM which can achieve this
consists of adding a charge $+2/3$ quark $Q$
whose left-handed and right-handed components are
both singlets under $S U (2)_L$.
The new quark $Q$ mixes with the standard charge $+2/3$ quarks,
and as a result $\Gamma_c$ could be reduced \cite{Ma}.
Isosinglet charge $+2/3$ quarks
have been considered in models
where the supersymmetric gauge group is extended \cite{BH}.
At the phenomenological level,
it could lead to \cite{BPR} a significant enhancement of
$D^0 - \bar{D}^0$ mixing,
detectable at  the next generation of experiments.
More drastically,
the charge $+2/3$ isosinglet quark might itself be
the 180 GeV quark observed at the Tevatron,
while the actual top quark remains hidden below  $M_W$
via fast $t\to cH^0$ decay induced by
large $c$-$Q$ and $t$-$Q$ mixings \cite{HH}.
As we shall see later, the latter scenario provides us with
a provocative possibility for {\it both} $R_b$ and $R_c$
to move in the right directions in a {\it substantial} way.

The minimal extension of
adding only one charge $+2/3$ isosinglet quark $Q$
leads to  new gauge invariant mass terms of the type
$\bar{Q}_L Q_R$ and $\bar{Q}_L u_{jR}$,
as well as additional Yukawa coupling terms $\bar{u}_{iL} Q_R$,
where $u_i$ denotes standard up-type quarks.
For the sake of simplicity, let us for now ignore the first generation
and set the standard KM mixing matrix to unity \cite{convention}.
One then has the charged current
\begin{equation}
(\bar c_L\ \bar t_L\ \bar Q_L)
          \left( \begin{array}{cc}
                    C_2 & 0 \\ -S_2S_3 & C_3 \\ +S_2C_3 & S_3
          \end{array} \right) \gamma_\mu
          \left( \begin{array}{r}
                    s_L \\ b_L
          \end{array} \right),
\end{equation}
where $S_i \equiv \sin\theta_i$, $C_i \equiv \cos\theta_i$.
The isospin part of the neutral current becomes
\begin{equation}
(\bar c_L\ \bar t_L\ \bar Q_L)
          \left[ \begin{array}{ccc}
                    C_2^2 & -S_2C_2S_3 & +S_2C_2C_3 \\
                   -S_2C_2S_3 & C_3^2 + S_2^2S_3^2 & C_2^2S_3C_3 \\
                  +S_2C_2C_3 & C_2^2S_3C_3 & S_2^2C_3^2 + S_3^2
          \end{array} \right] \gamma_\mu
          \left( \begin{array}{r}
                    c_L \\ t_L \\Q_L
          \end{array} \right).
\end{equation}
The only immediate change
accessible in $Z$ decay is
in the $Zc\bar c$ coupling,
\begin{equation}
 v_c = \sqrt{\rho}~ \left[t_3^c C_2^2
  - 2 Q_c \sin^2\bar{\theta}_W\right], \ \ \ \
 a_c  = \sqrt{\rho}~ t_3^c C_2^2,
\label{coupl}
\end{equation}
where $\sqrt\rho$ represents the non-trivial
wave function
renormalization of the 
$Z$ boson, and $\bar{\theta}_W$ is
the effective weak angle 
at the $Z$-pole. One finds
\begin{eqnarray}
\label{Rl}
R_l & \simeq & R_l^{\rm SM} \left(1 - 0.41 S_2^2
                            + 0.30 S_2^4\right),  \\
\label{Rb}
R_b & \simeq & R_b^{\rm SM}/\left(1 - 0.41 S_2^2
              + 0.30 S_2^4\right), \\
\label{Rc}
R_c & \simeq & R_c^{\rm SM} \left(1 - 2.41 S_2^2 + 1.75 S_2^4\right)
        /\left(1 - 0.41 S_2^2 + 0.30 S_2^4\right).
\end{eqnarray}
The SM expressions are given,
to very good approximation, as
\begin{eqnarray}
\label{smr}
R_l^{\rm SM} & = & (20.2 + 10.0\ \delta\rho + 6.3\;\ \delta V_b^t)
                         (1 + {{\alpha_S(m_Z)}/\pi} + \cdots),
 \nonumber \\
R_b^{\rm SM} & = & 0.220 - 0.01\ \delta\rho + 0.25\ \delta V_b^t, \\
R_c^{\rm SM} & = & 0.170 + 0.015\,\delta\rho - 0.055\,\delta V_b^t.
                                              \nonumber
\end{eqnarray}
where $\delta\rho$ denotes the deviation of $\rho$ from unity
which is mainly due to the $t$--$b$ splitting,
and $\delta V_b^t$ corresponds to the non-universal correction
to the $Zb\bar{b}$ vertex.
The leading Higgs dependence in $\delta \rho$ is logarithmic,
whereas in $\delta V_b^t$ the Higgs dependence is practically negligible.
Within the framework of the SM, $\delta \rho$ and $\delta V_b^t$, then
are given by ($x_i = m_i^2/m_Z^2$)
\begin{equation}
\delta\rho  \simeq
 (3\alpha/4\pi \sin^2{2\bar{\theta}_W})~x_t,
                                \ \ \ \
\delta V_b^t  \simeq  - (1.05 \alpha/ \pi) (x_t
 + 2.17~{\rm ln} x_t).
\end{equation}

We now extract the bound on $S_2^2$ from
$R_l^{\rm exp} = 20.788 \pm 0.032$,
and examine the implications for $R_b$ and $R_c$.
To accommodate a large mixing angle
we need a large value of $\alpha_S$,
as is evident from the expression of $R_l^{\rm SM}$ in eq. (\ref{smr}).
$R_b$ and $R_c$, on the other hand, are independent of
$\alpha_S$ for all practical purposes.
The combined average of LEP $+$ SLD is
$\alpha_S(m_Z) = 0.123 \pm 0.004 \pm 0.002$ \cite{lep}.
Since we use $R_l$ to extract $S_2^2$, we cannot use the
value of $\alpha_S(m_Z)$ derived from $R_l$ within the framework of SM.
However, various other independent
measurements of $\alpha_S$ at LEP
({\it e.g.} the comparison between 3-jet and 2-jet events, or, say,
from $\tau$-polarization measurement)
are consistent with each other.
We therefore choose two representative $\alpha_S(m_Z)$ values
\cite{webber}
as 0.123 and 0.126.
Since $R_l$ is almost flat with respect to variation of $m_t$,
we fix $m_t = 180$ GeV.
We take 70 GeV and 300 GeV as two representative values for $m_{H^0}$.
A low $m_{H^0}$ is slightly preferred to maximize the allowed $S_2^2$
and hence enhance the effects in $R_b$ and $R_c$.
The bounds on $S_2^2$ derived from the $2\sigma$ lower limit of $R_l$,
and the consequent changes in $\delta R_b$ and $\delta R_c$
using those angles, are displayed in Table \ref{table1}.

It is clear from Table \ref{table1} that,
although the shifts are in the right directions,
the discrepancies in $R_b$ and $R_c$ are
hard to make up in this minimal scenario.
What one really needs is a scenario
where $\Gamma_b$ is increased while $\Gamma_c$ is accordingly reduced,
such that $R_l$ remains consistent with experiment
-- a situation which could shift $R_b$ and $R_c$ in the right directions
by significant amounts.
However, {\it no minimal extension beyond the SM discussed so far in
the literature can do this job satisfactorily}.
In minimal supersymmetry,
light superpartners ($\sim$ 50--70 GeV) could
increase $\Gamma_b$ and push up $R_b$
by $\simeq 2\sigma$,
but $R_c$ remains practically untouched \cite{kane}.
This results in a lower $\alpha_S (m_Z) \simeq 0.118$
which is in consonance with lower energy measurements,
but not compatible with a simple supersymmetric Grand Unified Theory
 \cite{SGUTS}.
The distinctive feature of the isosinglet charge $+ 2/3$ quark scenario
is that $R_c$, which in most scenarios is hard to move,
can be brought down by $\simeq 0.4 \sigma$
by directly affecting $\Gamma_c$ \cite{Ma},
while the indirect effect on $R_b$ is also non-negligible
($\simeq 0.5 \sigma$ upward pull).
Note, however, that the $c$-$Q$ mixing angle $S_2$
has been singled out, while  $S_3$
is tacitly assumed to be smaller.
This is not particularly natural, for one might expect
the $t$-$Q$ mixing angle $S_3$ to be greater than $S_2$.
Allowing for large $S_3$, one could consider an intriguing effect
of  the so called
``light (hidden) top" scenario of ref.\cite{HH},
which we now elaborate.

With both $t$ and $Q$ entering the $Zb\bar b$ vertex correction,
the charged and neutral current couplings
that appear in the loop are modified through eqs. (2) and (3).
The impact can be summarized as an {\it effective} top mass,
\begin{equation}
m_t^2 \longrightarrow (m_t^{\rm eff.})^2
                     = m_t^2 + 2S_3^2m_t(m_Q - m_t)
                      + S_3^2(S_2^2 + S_3^2 - S_2^2S_3^2)(m_Q - m_t)^2.
\label{mteff}
\end{equation}
If one takes $m_t = 180$ GeV and $m_Q > m_t$,
this relation then dictates that $S_3^2$ has to be very small to avoid
aggravating the situation with $R_b$.
In the scenario of ref. \cite{HH}
(originally proposed in ref. \cite{MN}), however,
it is proposed that $m_t < M_W$ is possible because of
fast $t\to cH^0$ decay as compared to the standard $t\to bW^*$ mode,
which allows the top quark to evade earlier searches
by the CDF Collaboration.
It is then natural to take $m_Q = 180$ GeV to be
the heavy quark that is observed at the Tevatron.
This could work {\it only if}
both $S_2^2$ and $S_3^2$ are sizable \cite{HH}.
If such is the case, then $m_t^{\rm eff.}$
in the $Zb\bar b$ loop could be much smaller than 180 GeV.
It has been known for a long time that {\it  the larger $R_b$ value
favors a lighter $m_t$ than suggested from the global
electroweak fit} (and later, the Tevatron discovery).
We now have a mechanism to fit both demands,
hence it is worthwhile to redo our analysis.
We will come back to the issue of $R_l$ later,
and for now just concentrate on $R_b$ and $R_c$.

Let us illustrate how the mechanism would work.
First, $R_b^{\rm SM}$ in eq. (\ref{Rb})  should become
$R_b^{\rm SM}(m_t^{\rm eff})$,
with $m_t^{\rm eff}$ as given in eq. (\ref{mteff}).
Since $R_c$, in particular, may not be
as far away from its SM value as in eq. (1),
we assume that $R_b$ is shifted by $+3.7\sigma$,
while $R_c$ is shifted by $-1.4\sigma$,
such that $R_b + R_c$ is not drastically different from SM,
{\it i.e.}
 \begin{equation}
   R_b \simeq 0.2219\;\; (+3.7\sigma\ {\rm shift}), \;\;\;
   R_c \simeq 0.1616\;\; (-1.4\sigma\ {\rm shift}).
\label{rbrc}
\end{equation}
Thus, with $R_c/R_c^{\rm SM} \simeq 0.940$,
from eq. (\ref{Rc}) we find
\begin{equation}
S_2^2 = 0.0305,
\label{s2}
\end{equation}
which is larger than the values of $0.007$ -- $0.009$
given in Table I. Substituting into eq. (\ref{Rb}),
we find that $R_b^{\rm SM}(m_t^{\rm eff.}) = 0.219$,
which implies that  $m_t^{\rm eff.} \simeq 100$ GeV.
In the scenario of ref. \cite{HH},
we could, for example, take $m_t = 70$ GeV and $m_Q = 180$ GeV.
Solving eq. (\ref{mteff}) we get
\begin{equation}
S_3^2 \simeq 0.27,
\label{s3}
\end{equation}
which is larger than $S_2^2$.
Note that $t$ is still dominantly
the $SU(2)$ partner of the $b$ quark, which justifies our flavor label.
It may be noted , however, that with $C_2S_2S_3 \simeq 0.089$,
the $\bar c_L t_R H^0$ Yukawa coupling is
a factor of 3.5 weaker than in  ref. \cite{HH},
and if $m_{H^0} \gtrsim 60$ GeV, phase space suppression of
$t\rightarrow cH^0$ is itself too severe to allow it to dominate over
the standard $t\rightarrow bW^*$ mode.
We turn, however, to the issue of $R_l$,
the resolution of which results in a possible way out from
this problem as well.
Note that the present mechanism naturally does not affect
$\Gamma_d$ and $\Gamma_s$, and could be chosen not to
affect $\Gamma_u$.

It should be emphasized that,
within the present setup (minimal addition of $Q$),
$R_l$ cannot be accounted for.
The reason is because something similar to eq. (\ref{mteff})
would also enter into $\delta\rho$, making it smaller
than the SM value for $m_t = 180$ GeV.
To be more precise,
defining $\delta\rho =
(3\alpha/4\pi\sin^2{2\bar{\theta}_W})~\hat{T}$,
one finds \cite{ls}
\begin{equation}
 \hat{T}  =
x_t + S_3^2\left(-x_t + x_Q - C_3^2~f_{tQ}\right)
 + S_2^2\left(S_2^2 S_3^2 x_t + S_2^2 C_3^2 x_Q +
(2-S_2^2) S_3^2 C_3^2~f_{tQ} \right),
\label{rho2}
\end{equation}
where
$f_{ab} =  x_a + x_b + 2/(x_a^{-1} - x_b^{-1})\ln x_a/x_b$.
Using eqs. (\ref{s2}) and (\ref{s3}) and $m_t, m_Q = 70, 180$ GeV, one
obtains $\hat{T} = 1.14$, which should be compared with $\hat{T}^{\rm SM}
(m_t = 180) 
= 3.90$.
However, $\hat{T}$
is largely a measure of the accumulative effect of doublet splitting
in Nature, which enters into the $W$ and $Z$ boson two-point functions
(vacuum polarization).
In contrast, the effects that we discuss are for the
{\it flavor specific}
$Zc\bar c$ (tree level) and $Zb\bar b$
($t$ and $Q$ in loop) three-point functions \cite{Guido}.
Departing from the minimal addition of a singlet quark $Q$,
it is easy to conceive other sources of doublet splitting that
affect mainly the two-point functions and only marginally
the flavor specific three-point functions.
A convenient example is a second Higgs doublet
with $m_{H^+} > v$ but $m_{h^0} < M_W$,
where $h^0$ stands for lightest neutral (pseudo)scalar.
This gives a {\it positive} contribution
$\hat{T}^{h} = f_{+0}/3$, where `$+$' and `$0$'
stand for $m_{H^+}$ and $m_{h^0}$.
Numerically,
taking $m_{H^+}, m_{h^0} = 300, 60$ GeV,
one obtains $\hat{T}^{h} = 2.78$.
Hence, a sufficiently split second scalar doublet
could mimic the effect of the SM top-contribution to $\delta\rho$,
and help $R_l$ maintain its near-perfect agreement with observation.
As an extra bonus, $m_{h^0} < 60$ GeV becomes possible.
In fact, no realistic limit exists on the lighter neutral Higgs
in a two-doublet scenario.
Thus, a relatively light $h^0$ boson
(which is not directly produced in $e^+e^-$ collisions)
could help overcome the above mentioned phase space suppression
in the decay mode $t \to ch^0$, and
facilitate the hiding of the top below $M_W$ \cite{Hou}.
Note that a heavy $H^+$ boson makes little impact on
low energy observables such as $b\to s\gamma$ and $B^0$-$\bar B^0$
mixing.

We turn towards phenomenological consequences
and check whether one runs into conflict with other data.
In the conservative approach, because of the smallness of $S_2$
and the tacit assumption that $S_3$ is even smaller, there is practically
no observable consequences,
beyond the insufficient negative pull on $R_c$.
However, for the more provocative case,
because both $S_2$ and $S_3$ are quite sizable,
there is considerable impact on phenomenology \cite{HH},
especially those involving the top and singlet quarks.
First, from eq. (2) one sees that
$V_{cs}$ gets modulated by $C_2 \simeq 0.985$,
which is fully compatible with present errors.
Second, one finds
$V_{tb} \simeq C_3 \simeq 0.85$,
which may appear to be a bit small.
However,
in this scenario, it is the (dominantly) singlet quark $Q$ that is
``faking" the SM top quark at the Tevatron.
{}From eq. (2) one then finds that
$\vert V_{Qb}\vert \simeq S_3 \simeq 0.52,~
\vert V_{Qs}\vert \simeq S_2C_3 \simeq 0.15,$
which is in apparent conflict with recent studies
at the Tevatron that give
$\vert V_{\text{`}t\text{'}b}\vert = 0.97 \pm 0.15 \pm 0.07$ \cite{Vtb}.
On closer inspection, however,
what is actually measured is the ``$b$-content"
of top events.
If $m_Q \simeq 180$ GeV,
the leading decays are
$Q\to bW$, $sW$; $tH$, $tZ$; $cH$, $cZ$.
Using eqs. (2), (3), (\ref{s2}) and (\ref{s3}), we find their
relative weights 66.3\%, 5.4\%, 14.6\%, 7.3\%, 3.3\%, 3.0\%,
respectively.
Since $t\to cH^0$, $bW^*$ and $H^0 \to b\bar b$,
the  final state contains $b$ quarks whenever it contains $t$ or $H^0$.
Thus, the $b$ quark content of $Q$ decay is close to unity,
and $\vert V_{\text{`}t\text{'}b}\vert \simeq
\sqrt{0.663+0.146+0.073+0.033}
\simeq 0.96$, which is fully consistent with Tevatron results.

The implication for LEP-I is that, with our choices of $S_2$ and
$S_3$, $\Gamma(Z\to t\bar{c}+\bar{t}c) \sim 1$ MeV \cite{BH,HH},
but this is no easy task to measure.
We also note that the impact on $A_{FB}^f$ and $A_f$
are within errors.
In contrast, the consequence at LEP-II is dramatic.
The ``light (hidden) top" scenario demands that
$m_h \lesssim m_t \lesssim M_W < m_Q \simeq 180$ GeV.
Thus, unless the Tevatron could rule out
the scenario by detailed analysis below $M_W$ \cite{Hou,habermor}
(without assuming SM decay),
{\it toponium may show up soon at LEP-II}.
One could also observe open $t\bar t$,
and check that the decay rate is indeed faster than in SM.
At the same time, one should be able to discover a relatively
light Higgs boson in the decay products
of $t\bar t$ events.

Turning our attention to $D^0$--$\bar D^0$ mixing,
we note that it is expected to be small in the SM,
while the present scenario can induce a substantial effect.
The $Z$ mediated contribution gives \cite{BPR}
$\Delta m_D \simeq \sqrt{2} G_F f_D^2 B_D m_D \eta_{\text{QCD}}
S_1^2 S_2^2 /3 \sim 1 \times 10^{-7} S_1^2 S_2^2$ GeV,
hence it depends crucially on the size of $S_1$.
We have set $S_1 = 0$
from the outset, but it is in principle a free parameter,
just like $S_2$ and $S_3$.
Furthermore, it has been shown \cite{BPR,HH} that the
hierarchy $S_1 : S_2 : S_3 \sim m_u : m_c : m_t$
does not necessarily hold.
{}From the
experimental limit $\Delta m_D < 1.3 \times 10^{-13}$ GeV \cite{PDG},
we obtain the limits $\vert S_1\vert \lesssim 0.012,\ 0.006$ for
$S_2^2 = 0.008, \ 0.0305$, respectively.
Note that with such an $S_1$ value,
$B^0$--$\bar B^0$ mixing in the present scenario
can have both $t$ and $Q$
(with $m_t \lesssim M_W$ and $m_Q \simeq 180$ GeV)
quarks in the loop,
but can still be accounted for.


In summary,
we present the case of adding a charge $+2/3$
isosinglet quark as a possible solution to the so-called
$R_b - R_c$ problem.
In the conservative, minimal scenario where just one such quark
is added and $m_t$ is taken as $\sim 180$ GeV,
the precisely measured quantity $R_\ell$
can tolerate only a $-0.4\sigma$ pull on $R_c$, while
generating a $+0.5\sigma$ pull on $R_b$.
The direction is right, but the magnitude is insufficient.
However, in the ``light (hidden) top" scenario,
one could take $m_t < M_W$ and
identify the dominantly singlet quark as weighing 180 GeV.
One could then in principle allow
$\delta R_c \sim -1.4\sigma$ while inducing
$\delta R_b \sim +3.7\sigma$.
The push or pull comes about because of $c$-$Q$ mixing,
just like in the conservative case,
but in addition, $R_b$ is raised due to a lower effective top mass.
The common, key ingredient
is to have rather sizable $c$-$Q$ and $t$-$Q$ mixings.
It is, however, necessary to add a sufficiently split
scalar doublet to simulate the effect of a standard heavy top
on the oblique parameters, as inferred by the global fit.
At least one neutral Higgs boson should be rather light
such that fast $t\rightarrow ch^0$ decay would not be
hindered by phase space.
The existence of toponium and light Higgs bosons
are dramatic consequences that can be tested immediately at LEP-II,
where the model could be fully confirmed or ruled out.
Unlike MSSM solutions to $R_b$ (but not $R_c$) problem,
the present scenario is rather {\it ad hoc, i.e.}
tailored to the $R_b$--$R_c$ problem.
But that is also an advantage,
for its effects do not show up strongly elsewhere
except when involving heavy charge $+2/3$ quarks.
If new particles are found at LEP II,
one should strive to distinguish between the
light chargino/stop option
of minimal supersymmetric Standard Model,
{\it vs.} the light top/Higgs scenario with exotic charge
$+2/3$ isosinglet quarks.

\acknowledgments
GB thanks Suchandra Dutta for private communication regarding electroweak
fit and D. Choudhury for discussions.
The work of GCB was supported in part by Science Project No.
SCI-CT91-0729 and E.C Contract No. CHRX-CT93-0132.
WSH is supported in part by grant NSC 85-2112-M-002-011
of the Republic of China. He also wishes to thank the hospitality of the
CERN Theory Division.

\begin{table}
\caption[] {
Bounds on $S_2^2$ from 2$\sigma$ lower limit of $R_l$
for a fixed $m_t = 180$ GeV, but for different values of $m_H$ and
$\alpha_S(m_Z)$, are displayed.
$R_l^{\rm SM}$ for various inputs have been calculated \cite{chumki}
using the ZFITTER version 4.9.
Also shown are the corresponding $\delta R_b$ and $\delta R_c$.
For $\alpha_S(m_Z) = 0.123$ and $m_H = 300$ GeV, $R_l$
lies below the 2$\sigma$ lower limit of $R_l^{\rm SM}$.
}
\begin{tabular}{cccccccd}
  $m_H$ & & $\alpha_S(m_Z) = 0.123$ &
                & & $\alpha_S(m_Z) = 0.126$ & \\ 
  (GeV) & $S_2^2$ & $\delta R_b$ & $\delta R_c$
                & $S_2^2$ & $\delta R_b$ & $\delta R_c$ \\ \hline
  70  &  0.007 & 0.0006 & $-$0.0024
                & 0.009 & 0.0008 & $-$0.0032 \\ \hline
  300 &  --    & --     &  --     & 0.008 & 0.0007 & $-$0.0026 \\
\end{tabular}
\label{table1}
\end{table}


\begin{references}
\bibitem{SGUTS} Ignoring possible threshold effects
and contribution from higher dimensional operators,
$\alpha_S (m_Z)$ does not descend below $0.125$ in such a theory.
\end{references}
\end{document}